\documentclass[12pt,preprint]{aastex}

\newcommand{\kms}{\>{\rm km}\,{\rm s}^{-1}}

\newcommand{\msun}{\>{\rm M_{\odot}}}
\newcommand{\lsun}{\>{\rm L_{\odot}}}

\newcommand{\bdm}{\begin{displaymath}}
\newcommand{\edm}{\end{displaymath}}
\newcommand{\beq}{\begin{equation}}
\newcommand{\eeq}{\end{equation}}
\newcommand{\bit}{\begin{itemize}}
\newcommand{\eit}{\end{itemize}}
\newcommand{\ben}{\begin{enumerate}}
\newcommand{\een}{\end{enumerate}}
\newcommand{\bfi}{\begin{figure}[htb]}
\newcommand{\bpfi}{\begin{figure}[p]}

\newcommand{\name}{J1000+0234}
\newcommand{\names}{J1000+0234 ~}

\slugcomment{draft version}

\shorttitle{Molecular Gas in a z=4.5 Submillimeter Galaxy}
\shortauthors{Schinnerer et al.}

\begin{document}


\title{Molecular Gas in a Submillimeter Galaxy at z=4.5: \\
Evidence for a Major Merger at 1 Billion Years after the Big
Bang\altaffilmark{1} }


\author{E. Schinnerer\altaffilmark{2},
C.L. Carilli\altaffilmark{3},
P. Capak\altaffilmark{4,5},
A. Martinez-Sansigre\altaffilmark{2},
N.Z. Scoville\altaffilmark{5},
V. Smol\v{c}i\'{c}\altaffilmark{5},
Y. Taniguchi\altaffilmark{6}
M.S. Yun\altaffilmark{7},
F. Bertoldi\altaffilmark{8}
O. Le Fevre\altaffilmark{9},
L. de Ravel\altaffilmark{9}}

\altaffiltext{1}{Based on observations carried out with the IRAM
  Plateau de Bure Interferometer. IRAM is supported by INSU/CNRS
  (France), MPG (Germany) and IGN (Spain).}

\altaffiltext{2}{Max-Planck-Institut f\"ur Astronomie, K\"onigstuhl 17,
    D-69117 Heidelberg, Germany}
\altaffiltext{3}{National Radio Astronomy Observatory, P.O. Box O, Socorro, 
  NM, 87801, U.S.A.}
\altaffiltext{4}{Spitzer Science Center, 314-6 CalTech, Pasadena, CA, 91125, 
  U.S.A.}
\altaffiltext{5}{California Institute of Technology, MS 105-24, Pasadena, 
  CA, 91125, U.S.A.}
\altaffiltext{6}{Research Center for Space and Cosmic Evolution, 
 Ehime University, 2-5 Bunkyo-cho, Matsuyama 790-8577, Japan}
\altaffiltext{7}{Astronomy Department, University of Massachusetts, Amherst, 
 MA, 01003, U.S.A.}
\altaffiltext{8}{Argelander Institut f\"ur Astronomie, Universit\"at Bonn,
 Auf dem H\"ugel 71, D-53121 Bonn, Germany}
\altaffiltext{9}{Laboratoire d'Astrophysique de Marseille, 
 BP 8, Traverse du Siphon, 13012 Marseille Cedex 12, France }

\begin{abstract}
  We report the detection of CO molecular line emission in the z=4.5
  millimeter-detected galaxy COSMOS\_J100054+023436 (hereafter:
  \name) using the IRAM Plateau de Bure interferometer (PdBI) and
  NRAO's Very Large Array (VLA). The $^{12}$CO(4-3) line as observed
  with PdBI has a full line width of $\sim1000\,\kms$, an integrated
  line flux of 0.66 Jy\,$\kms$, and a CO luminosity of
  3.2$\times10^{10}\,\lsun$. Comparison to the 3.3$\sigma$ detection
  of the CO(2-1) line emission with the VLA suggests that the molecular gas
  is likely thermalized to the J=4-3 transition level. The corresponding
  molecular gas mass is $\rm 2.6\times10^{10}\,\msun$ assuming an
  ULIRG-like conversion factor. From the spatial offset of the red- and
  blue-shifted line peaks and the line width a dynamical mass of $\rm
  1.1\times10^{11}\,\msun$ is estimated assuming a merging scenario.
  The molecular gas distribution coincides with the rest-frame
  optical and radio position of the object while being offset by 0.5''
  from the previously detected Ly$\alpha$ emission. \names exhibits
  very typical properties for lower redshift (z$\sim$2) sub-millimeter
  galaxies (SMGs) and thus is very likely one of the long sought after
  high redshift (z$>$4) objects of this population. The large CO(4-3)
  line width taken together with its highly disturbed rest-frame UV
  geometry suggest an ongoing major merger about a billion years after
  the Big Bang. Given its large star formation rate (SFR) of
  $>1000\msun\,yr^{-1}$ and molecular gas content this object could be the
  precursor of a 'red-and-dead' elliptical observed at a redshift of
  z=2.
\end{abstract}



\section{Introduction}

Wide field blank sky surveys at millimeter and sub-millimeter
wavelengths have established a population of active star forming
galaxies at high redshift \cite[e.g.][]{bla02}. These sources or
so-called submillimeter galaxies (SMGs) dominate the (sub-)millimeter
background, and represent (50-75)\% of the star formation at high
redshift causing a significant revision to the optically derived star
formation history of the Universe \citep{sma97,hug98,bar98,ber00}.

The bulk of this population lies at redshifts below z=3, with
far-infrared (FIR) luminosities of $\ge10^{13}$L$_\sun$, which (if
powered by star-formation) imply star-formation rates in excess of
1000\,M$_\sun$ yr$^{-1}$. This is sufficient to build up the stellar
mass of a giant elliptical galaxy in about 1 Gyr \citep{cha05}. At
redshifts z$\le$3 about 100 SMGs are now known, many of which
have been studied in detail, including CO observations that indicate
that they are very massive systems \citep{gre05}. Recent high
resolution studies of the molecular gas in z$\sim$2-3 SMGs
\citep{tac06,tac08} showed that the star forming regions are fairly
compact and that the SMGs resemble ``scaled-up and more gas-rich
versions'' of the local Ultra-Luminous Infra-Red Galaxies
\citep[ULIRGs; e.g.][]{dow98}. Due to their derived central
densities, which are close to those of ellipticals or massive bulges,
they appear to form stars at the maximal rate over very short time
scales (''maximum starburst''). Thus the SMGs phase appears to last
for about 100\,Myr \citep{tac08}. These massive starburst galaxies
place tight constraints on galaxy formation models
\citep[e.g.][]{bau05}. One major focus of current and future (sub-)mm
surveys is to identify these massive starbursts at z$>$3, a redshift
range for which SMGs could place tight constraints on hierarchical
merger vs. monolithic collapse models in galaxy formation scenarios.

The discovery of such an SMG above a redshift of z=4 indicates that
massive galaxy formation is already well under-way when the universe
is only 1.5 Gyr old. The recently discovered object \object{\name}
originally selected as a V band drop-out with a weak radio counterpart
\citep{car08} and with a stellar mass of $\rm M_*>10^{10}\,\msun$, a
young starburst age of 2-8\,Myr and a star formation rate of SFR
$>1000\,\msun yr^{-1}$ is a candidate for such a z$>$4 SMG, as it was
detected in its FIR continuum \citep{cap08}. Throughout this paper we
assume a standard concordance cosmology ($\rm H_{\circ}=70,
\Omega_M=0.3, \Omega_{\Lambda}=0.7$).

\section{Observations}

The bolometer camera MAMBO-2 at the IRAM 30m telescope was used in
on-off mode in December 2007 and January 2008 to measure the 1.2\,mm
flux density of \name. The weather conditions during the observations
were good, the reduction was performed using MOPSIC (written by R.
Zylka, IRAM). The achieved rms for a total on-sky integration of
70\,min is 0.67\,mJy. \names was detected at a 5$\sigma$ level with a
total 1.2mm continuum flux density of 3.4\,mJy.

The $^{12}$CO(4-3) line tracing cold molecular gas in
\names was observed with the IRAM Plateau de Bure interferometer
(PdBI) between January and April 2008 in the special C and C
configurations. Both receivers were tuned to the redshifted line
frequency of 83.3257\,GHz covering a total bandwidth of $\sim$
0.9\,GHz. For calibration and mapping, we used the standard IRAM
GILDAS software packages CLIC and MAPPING \citep{gui00}. The quasar
1005+058 was used for atmospheric calibration while standard
calibrators were used for flux calibration. The 12\,hr of total
integration time result in a data cube of 2.3''$\times$1.9''
resolution (using robust weighting) with an rms of $\rm
0.42\,mJy\,beam^{-1}$ for 20\,MHz ($\sim72\,\kms$) wide channels.

We used the VLA in D configuration to observe the
$^{12}$CO(2-1) line in June and July 2008 for a total of 21\,hr. The
two 50~MHz wide correlator units (each corresponding to a velocity
width of $\sim360\,\kms$) were tuned to 41.6352\,GHz (on-line) and
41.7351\,GHz (off-line). The observations were done in the
fast-switching mode using the nearby quasar 1018+055, at a distance of
5.3$^{\circ}$ from our source, for atmospheric amplitude and phase
calibration. 20 antennas were available during the observations and
the phase stability was typically $\sim10^{\circ}$. The final images
using natural weighting have a resolution of 1.9''$\times$1.4'' and an
rms of 49\,$\rm \mu Jy\,beam^{-1}$.

All observed and derived properties of \names are summarized in Tab.
\ref{tab:prop}.

\section{The Molecular Gas Properties} 

\begin{figure}
\includegraphics[angle=-90.,scale=0.7]{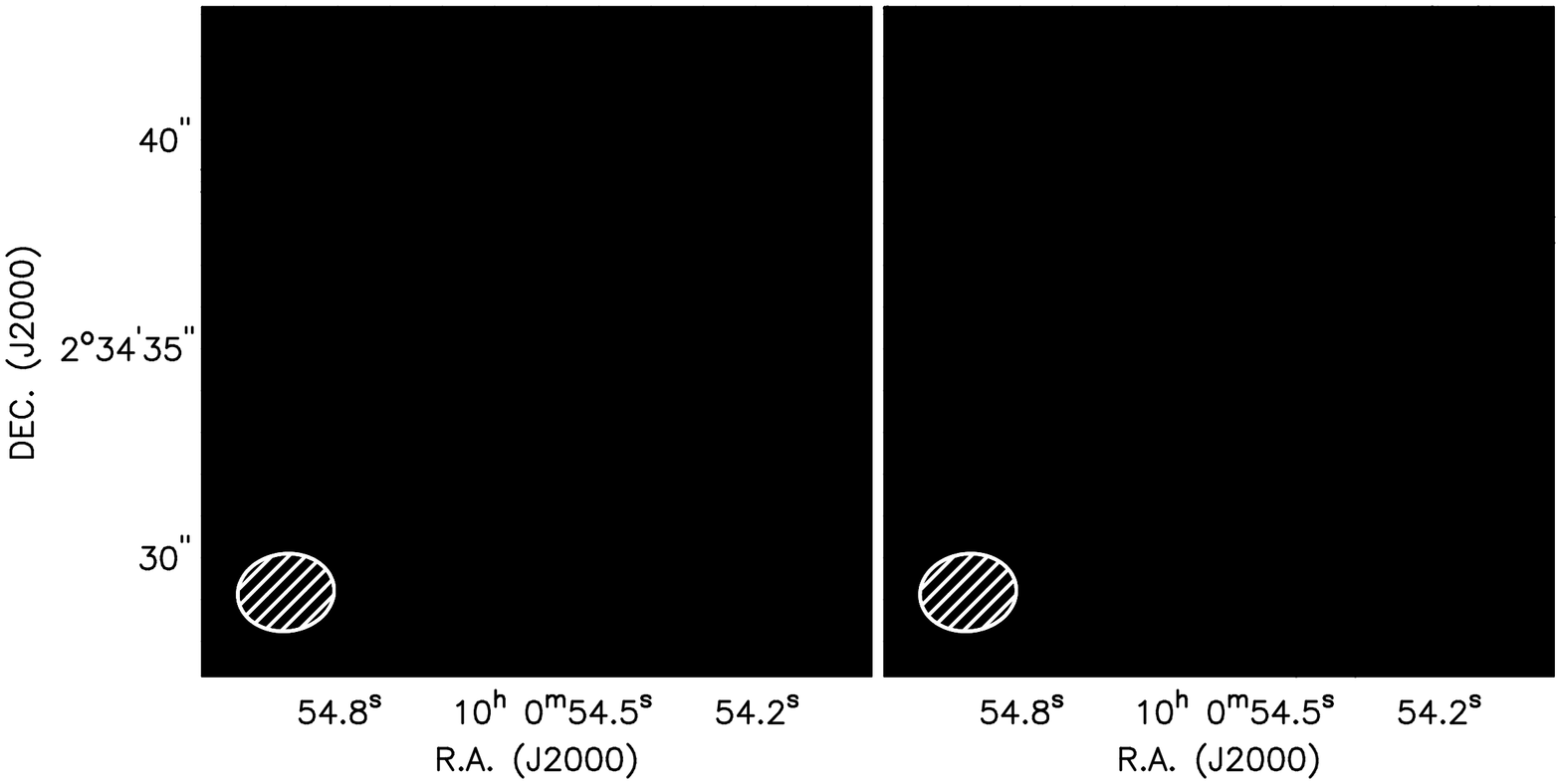}
\caption{CO(4-3) line emission and corresponding continuum in \names
  as observed with the PdBI, shown on the same color scale, but with
  different contour levels. Continuum emission from \names is detected
  at the 2$\sigma$ level, while the emission from the CO(4-3) line
  shows a solid detection at 5.5$\sigma$. The integrated CO(4-3) line
  emission ({\it left}) is centered at a velocity of +500\,$\kms$
  relative to the observed frequency of 83.3257\,GHz. Contours start
  at 2$\sigma$ in steps of 1$\sigma$ with 1$\sigma$ = 0.12\,$\rm
  Jy\,beam^{-1}\,\kms$. The integrated continuum ({\it right}) has the
  same contour steps, however 1$\sigma$ corresponds to 0.08 mJy/beam.
  The cross indicates the position of the CO(4-3) peak. The beam is
  shown in the lower left corner. \label{fig:co}}
\end{figure}

The CO(4-3) line was detected at a 5.5$\sigma$ level (Fig.
\ref{fig:co}) in a 280\,MHz wide channel, which corresponds to a full
line width at zero intensity (FWZI) of $\sim1000\,\kms$, i.e. with an
integrated flux of 0.66\,Jy\,$\rm \kms$. The peak position of the
CO(4-3) emission (see Fig. \ref{fig:comp}) agrees very well within the
positional uncertainties with the rest-frame near-IR and radio
position derived by \cite{cap08}. The line emission of CO(4-3) is
centered at an observed frequency of 83.1857\,GHz, which corresponds
to a redshift of z=4.5423 (assuming a rest frequency of
461.040768\,GHz for $^{12}$CO(4-3)). Given the large line width this
is in agreement with the optically derived redshift of
z=4.547$\pm$0.002 \citep{cap08}. The CO(4-3) emission coincides with
the rest-frame optical counterpart (as traced by the IRAC data) of
\names (Fig. \ref{fig:comp}). Averaging the remaining channels blue-
and redwards of the detected line emission results in a line-free
image of 620\,MHz width ($\sim2200\,\kms$) with a tentative
2.2$\sigma$ detection of the continuum at the rest-frame optical/radio
position of the source (Fig.  \ref{fig:co}).

The VLA observations with a channel width of $\sim360\,\kms$ are
centered at the middle of the line detected in CO(4-3). CO(2-1) line
emission is detected at the 3.3$\sigma$ level and coincides spatially
with the CO(4-3) emission within its positional uncertainties of
FWHM$_{beam}$/(S/N)$\sim$0.5''. Only about a third of the total line width
was covered during the VLA observations, so the detected line flux of
0.057\,Jy\,$\kms$ could be higher by a factor of $\sim$3, taking the missed
blue- and redshifted emission into account by assuming a box-car line
shape. The ratio of the CO(4-3) line flux of 0.66\,Jy\,$\kms$ to the
thus corrected CO(2-1) line flux of $\sim$0.16\,Jy\,$\kms$ is about 4,
implying that the molecular gas is still thermalized at the
J=4-3 transition, as the flux still increases with $\nu^2$ at this
transition.

Continuum emission from \names was detected at
1.2\,mm at (3.4$\pm$0.67)\,mJy, while the off-line
channels in the interferometric observations resulted in a 3$\sigma$
upper limit of 150$\mu$Jy at 41.7\,GHz (VLA) and a tentative
2.2$\sigma$ detection of (0.17$\pm$0.08)mJy at 83.3\,GHz (PdBI). All
these values are consistent with the SED fits presented by
\cite{cap08}\footnote{The scaling of the SEDs presented in Fig. 4 of
\cite{cap08} is overestimated by a factor
of 10$^9$ (P. Capak, priv. comm.).} and a dust temperature of 30 -
50 K, typically found in z$\sim$2 SMGs \citep{pop06,kov06}.

\section{Masses and Typical SMG Properties}

As the molecular gas appears to be thermalized at least up to the CO(4-3)
transition as inferred above, its line flux can be used to estimate
the CO luminosity [setting $\rm L'_{CO}(J=4-3)=L'_{CO}(J=1-0)$] and
hence the molecular gas mass present in \name. To derive the CO
luminosity $\rm L'_{CO}$ and molecular gas mass $\rm M_{gas}$ we use
equations 3 and 4 from \cite{sol05} with a conversion factor for $\rm
L'_{CO}$ to $\rm M_{gas}$ of $\alpha$=0.8\,$\rm
M_{\odot}(K\,\kms\,pc^2)^{-1}$ derived for local ULIRGs \citep{dow98}
and used for high redshift objects \citep{sol05}. The observed line
flux of $\rm S_{CO(4-3)}$=0.66\,Jy\,$\kms$ corresponds to a CO
luminosity of $\rm L'_{CO}\sim3.2\times10^{10}\,L_{\odot}$ or a
molecular gas mass of $\rm M_{gas}\sim2.6\times10^{10}\,\msun$.

We imaged the red and blue ($\approx$ 500\,$\kms$ wide) halves of the
line emission separately, resulting in a 4.7$\sigma$ (red) and 3.8$\sigma$
(blue) emission peak, respectively. The two peaks (Fig.
\ref{fig:velo}) show a spatial shift of $\sim$ 1'' roughly from
northwest to southeast (with a positional uncertainty of $\sim$0.5'').
This offset corresponds to $\sim$ 6.6\,kpc at the redshift of \name,
suggesting that the CO emission might be fairly extended. Using this
spatial offset between the red- and blue-shifted half of the line
emission and the line width of $\sim$1000\,$\kms$, we can estimate the
dynamical mass of \name. We use the relation between dynamical mass
$M_{dyn}$, velocity width $\rm \Delta\,v_{FWHM}=v_{rot}(r)\times
sin\,{\it i}/2.4$ and radial extent $r$ of the CO emitting region:
$\rm M_{dyn}\times sin^2{\it i} = 4\times10^4\,r\,\Delta\,v^2_{FWHM}$
\citep{ner03}. As the system is likely merging (as discussed below) we
include also a factor of 2 for a merging system assuming that the gas
has already settled into the new potential \citep{gen03}. The
estimated dynamical mass of \names is about 1$\times$10$^{11}\msun$
for values of $\rm v_{rot}(r)\times sin\,{\it i} \sim 1000\,\kms$ and
a radius $r$ of 3.3\,kpc. Thus \names has a gas fraction of $\rm
M_{gas}/M_{dyn} \sim 25\%$, typical for SMGs \citep{gre05,tac06}. We
would like to caution that the derived offset of $\sim$1'' has
significant uncertainties and thus the derived dynamical mass could be
overestimated if the separation were smaller.

The properties of \names are very similar to the global properties
derived for z$\sim$2 SMGs in terms of molecular gas mass ($\rm
<M_{gas}> \sim 3\times10^{10}\msun$), extent ($\rm r_{CO} \sim
2\,kpc$), CO line width ($\rm <FWHM_{CO}> \sim 780\,\kms$), dynamical
mass ($\rm <M_{dyn}> \sim 1.2\times10^{11}\,\msun$) and molecular gas
mass fraction ($\sim$ 25\%) \citep{gre05}. Its estimated current gas
consumption time scale of $\rm \tau=M_{gas}/SFR \approx 30\,Myr$
assuming our gas mass and a SFR of $\rm \ge 1000\,\msun\,yr^{-1}$
\citep{cap08} is similar to the gas depletion rates of $\sim$ 100\,Myr
found for the z$\sim$2 SMGs \citep{tac08}. Taking together the stellar
mass of $\rm \ge 10^{10}\,\msun$ produced in a recent burst $\sim$
7\,Myr ago \citep{cap08} and the available gas reservoir and gas
consumption time scale derived here, a significant fraction of the
stellar mass of a massive elliptical could be produced on a relatively
short time scale. Assuming that these recently formed stars will be
passively evolving since then, \names could turn into a 'red-and-dead'
elliptical -- like those that are found at z=2 when about 2\,Gyr have
passed since z=4.5.

\section{Evidence for an Ongoing Major Merger}

\begin{figure}
\includegraphics[angle=-90.,scale=0.7]{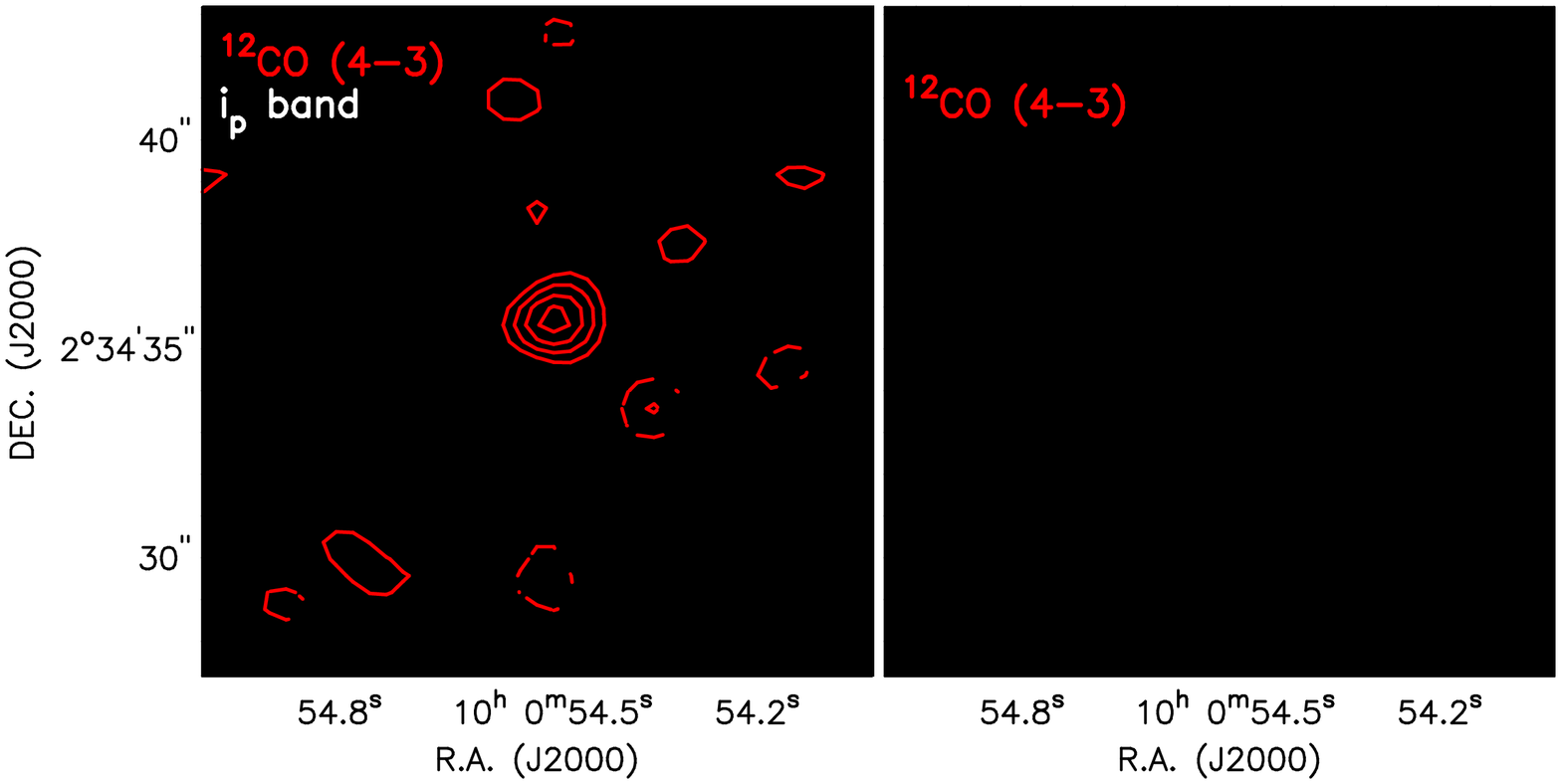}
\caption{Location of the molecular gas relative to other tracers.
  {\it Left:} The CO(4-3) line intensity overlaid in contours (same as
  in Fig. \ref{fig:co}) on the Subaru $i_p$ band image, showing an
  offset from the bright blue continuum rim.  {\it Right:} VLA
  1.4\,GHz continuum contours \citep[65\% to 95\% of the maximum of
  45$\mu$Jy;][]{sch07} overlaid onto the CO(4-3) map showing an
  excellent correspondence between the molecular gas and the radio
  continuum, which is presumably tracing star
  formation activity.  \label{fig:comp}}
\end{figure}

\begin{figure}
\includegraphics[angle=-90.,scale=0.9]{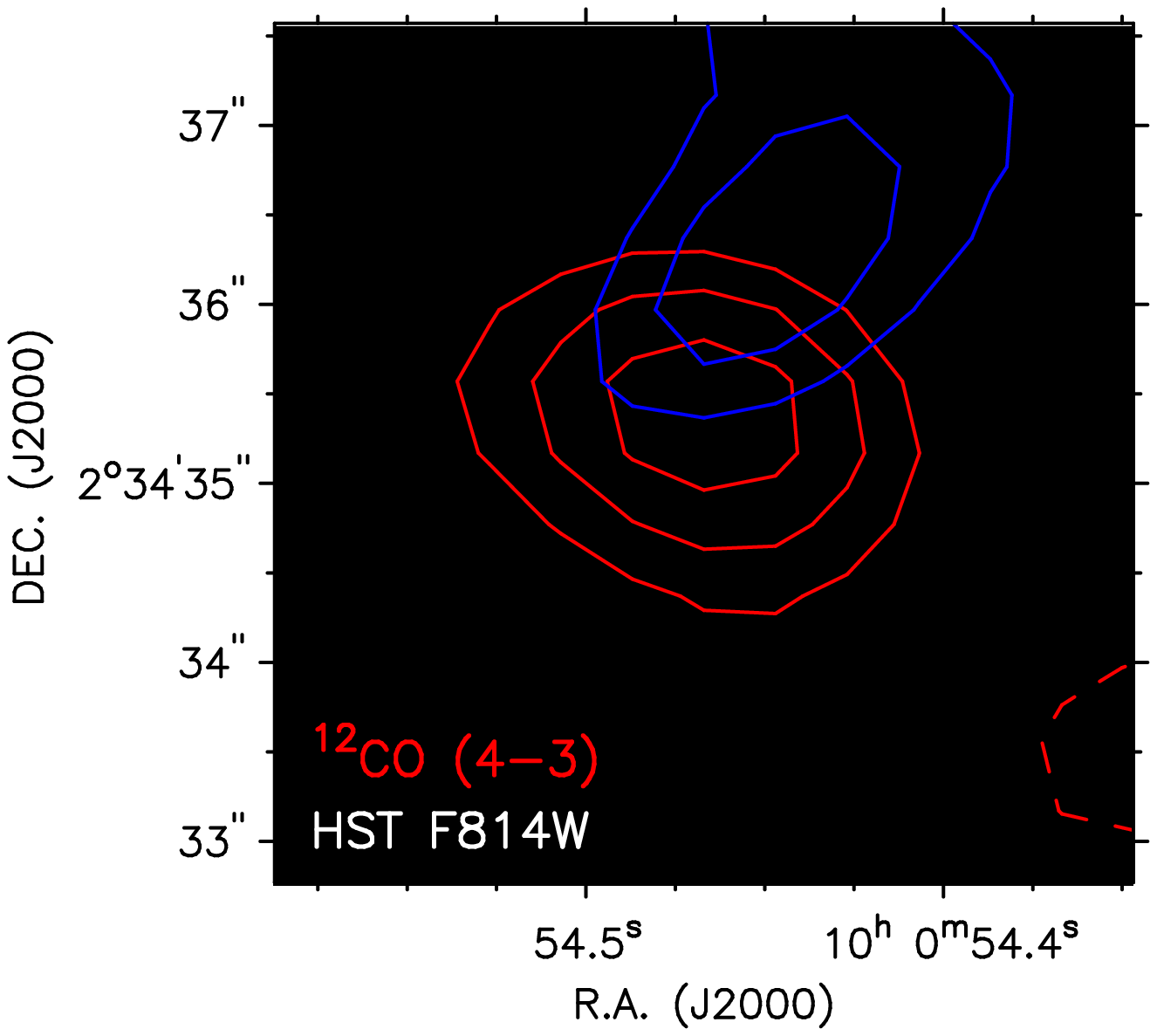}
\caption{Integrated red- and blueshifted CO(4-3) line emission in
  contours overlaid on the adaptively smoothed HST ACS/F814W image of
  \name. A motion from southeast to northwest is apparent with an
  offset of $\sim$1'' between the peaks of the red- and blueshifted
  emission.  The contours start at 2$\sigma$ in steps of 1$\sigma$
  with 1$\sigma$ = 0.085\,$\rm Jy\,beam^{-1}\,\kms$ and 0.084\,$\rm
  Jy\,beam^{-1}\,\kms$ for the red- and blueshifted line emission,
  respectively. The object $\sim$ 0.5'' west of the CO emission is a
  foreground galaxy at z=1.41 \citep{cap08}. \label{fig:velo}}
\end{figure}

The molecular gas emission arises from a region that is highly
obscured at the observed optical wavelengths \citep[see Fig. 1
of][Fig. \ref{fig:comp} and \ref{fig:velo}]{cap08}. However, it is
coincident with the position of the rest-frame optical and radio
emission (Fig. \ref{fig:comp}). The HST ACS image of \names
\citep{cap08} was adaptively smoothed \citep{sco00} to enhance the
structure present in the blue continuum covered by the F814W filter
(Fig. \ref{fig:velo}). The molecular gas appears to lie next to bright
blue continuum emission that is at its eastern side. Ly$\alpha$
emission has also been associated with this bright rim in blue
continuum \citep{cap08}. The motion of the gas is roughly along a
region of higher extinction (compared to the blue western component)
running roughly from north to south and shows a larger spatial extent
than the blue continuum emission. Note the object west of the gas
emission is a foreground galaxy at z=1.41. As already mentioned by
\cite{cap08} this geometry is very reminiscent of an ongoing merger.
As the redshift of the molecular gas is very close to the redshift
derived from the Ly$\alpha$ line (z=4.542 vs.  z=4.547), it is very
likely that both components belong to the same (highly disturbed)
system. Both the large FWZI of the CO(4-3) line and the appearance of
\names in Ly$\alpha$ make an ongoing merger scenario very likely. We
derive a conservative estimate of the merging time for \names of
significantly less than a billion years using the prescription of
\cite{kit08} for close galaxy pairs with velocities of $v <
3000\,\kms$. This time scale is consistent with having a relaxed system
at a redshift of z=2.

The expected number of z=4-5 SMGs from the model of \cite{bau05}
yields of the order of 10-20 such sources within the 2\,deg$^2$ COSMOS
field. As \names was originally identified as a Lyman Break Galaxy via
its V-band drop \citep{cap08,car08}, this suggests that UV emission
can be detectable from ongoing major mergers at these redshifts. We
have identified 4 additional V-band dropout LBGs with 4$\sigma$ radio
detections \citep{car08} that could also be SMGs at this redshift. An
inspection of their optical appearance in the HST ACS data shows that
two of them also have a disturbed geometry. This suggests that there
could be more objects sharing properties of both the SMG and LBG
population at z$>$4 and \names might therefore provide an important
link between the SMG and LBG populations at high redshifts.

The case of \names shows that (at least some) z$>$4 SMGs can have very
similar properties to their well-studied low redshift (z$\sim$2)
counterparts, that major mergers with very large SFRs are likely
present at z=4.5 and that these systems might be the precursors of
elliptical galaxies found at the red sequence at a redshift of z=2.

\acknowledgments

ES would like to thank J.M. Winter and R. Neri for their great help
during the visit of IRAM, Grenoble. ES thanks A. Wei\ss ~and E.F. Bell
for helpful discussions. ES acknowledges the hospitality of the Aspen
Center for Physics where the manuscript was prepared. CC thanks the
Max-Planck-Gesellschaft and the Humboldt-Stiftung for support through
the Max-Planck-Forschungspreis. The National Radio Astronomy
Observatory (NRAO) is operated by Associated Universities, Inc., under
cooperative agreement with the National Science Foundation.

{\it Facilities:} \facility{IRAM (PdBI,30M), NRAO (VLA)}

\clearpage

\begin{deluxetable}{lrl}
\tablecaption{Properties of \name \label{tab:prop}}
\tablewidth{0pt}
\tablehead{
\colhead{Property} & 
\colhead{Value}  &  
\colhead{Comment} } 
\startdata
$\rm S_{250\,GHz}$ [mJy]  & 3.4$\pm$0.67 & MAMBO\\ 
$\rm S_{83\,GHz}$ [mJy] & 0.2$\pm$0.09 & off-line channels ($\sim$2$\sigma$, PdBI)\\
$\rm S_{42\,GHz}$ [mJy] & $\le$0.15 & off-line channel ($\le$3$\sigma$, VLA)\\ \hline
$\rm z_{CO(4-3)}$ & 4.542 & center of CO(4-3) line\\
$\rm FWZI_{CO(4-3)}$ [$\kms$] & $\sim$ 1000  & CO(4-3) line\\ \hline
$\rm S_{CO(4-3)}$ [Jy\,$\kms$] & 0.66$\pm$0.12  & 5.5$\sigma$ detection (PdBI)\\
$\rm S_{CO(2-1)}$ [Jy\,$\kms$] & 0.057$\pm$0.017$^{\dagger}$ & 3.3$\sigma$ detection$^*$ (VLA) \\
$\rm R.A._{CO(4-3)}$ (2000)& 10:00:54.484 & line peak \\
$\rm Dec._{CO(4-3)}$ (2000) & +02:34:35.73 & line peak\\ \hline
$\rm L'_{CO} [L_{\odot}]$ & 3.2$\times$10$^{10}$ & from CO(4-3) \\
$\rm M_{gas} [M_{\odot}]$ & 2.6$\times$10$^{10}$ & from CO(4-3) \\
$\rm M_{dyn}\times sin^2{\it i} [M_{\odot}]$ & 1.3$\times$10$^{11}$ & using FWZI \& r=0.5''\\
\enddata 
\tablecomments{Some properties measured from our PdBI and VLA data and 
derived using the equations given by \cite{sol05} and \cite{ner03}.
\\
$^{\dagger}$ observed line flux not corrected for small bandwidth (see text for details)\\
$^*$ available bandpass corresponds to $\Delta$v$\approx$360 $\kms$, i.e. a third of the total line width
}
\end{deluxetable}

\clearpage

\clearpage

\clearpage

\end{document}